\documentclass[useamsfonts]{pasj00}
\usepackage{graphicx}
\begin{document}
\hyphenation{Schwarz-schild}

\SetRunningHead{J.~Hor\'ak et al.}{NBOs in non-linear resonance}
\Received{2004/06/01}
\Accepted{2004/07/27}

\title{Of NBOs and kHz QPOs: a low-frequency modulation
\\ in resonant oscillations of relativistic accretion disks}

\author{J.~\textsc{Hor\'ak},\altaffilmark{1}
M.\,A.~\textsc{Abramowicz},\altaffilmark{2}
V.~\textsc{Karas},\altaffilmark{3,1}
W.~\textsc{Klu{\'z}niak},\altaffilmark{4,5}}
\affil{$^1$~Faculty of Mathematics and Physics, Charles University,
   V~Hole{\v s}ovi\v{c}k{\'a}ch 2, CZ-180\,00 Prague, Czech Republic}
\affil{$^2$~Department of Theoretical Physics, Chalmers University
 and G\"oteborg University, S-412\,96 G{\"o}teborg, Sweden}
\affil{$^3$~Astronomical Institute, Academy of Sciences,
   Bo\v{c}n\'{\i}~II, CZ-141\,31 Prague, Czech Republic}
\affil{$^4$~Institute of Astronomy, Zielona G{\'o}ra University, Lubuska 2,
    P-65\,265 Zielona G{\'o}ra, Poland}
\affil{$^5$~Copernicus Astronomical Center, ul. Bartycka 18, P-00\,716,
 Warszawa, Poland}
\maketitle
\begin{abstract}
The origin of quasi periodic modulations of flux in the kilohertz range
(kHz QPOs), observed in low-mass X-ray binaries, is usually assumed to
be physically distinct from that of the ``normal branch oscillations''
(NBOs) in the Z-sources.  We show that a low-frequency modulation of the
kHz QPOs is a natural consequence of the non-linear relativistic
resonance suggested previously to explain the properties of the
high-frequency twin peaks. The theoretical results discussed here are
reminiscent of the $6$~Hz variations of frequency and amplitude of the
kHz QPOs reported by Yu, van der Klis and Jonker (2001).
\end{abstract}

\section{Introduction}
\label{intro}
Power density spectra of the X-ray flux of low-mass X-ray binaries have 
a rich phenomenology almost entirely lacking a physical explanation
(\cite{vdk1}, \yearcite{vdk00}).  Twin kHz QPOs in neutron star sources,
and their kHz counterpart in black-hole systems, have attracted the most
attention because their frequency is comparable to the orbital
frequencies in the innermost dynamically stable regions of the accretion
flow \citep{kmw,vdk2}.  {\it Ad hoc} models of QPOs have not withstood
the test of time, but a more fundamental approach of studying the
eigenmodes of accretion disks remains promising \citep{wag,kato}. It
seems necessary to include non-linear effects to be in agreement with
observations \citep{tokyo}. Here, we wish to point out that even the
simplest non-linear equations of motion have a rich structure which is
yet to be explored fully in the context of QPOs.

It is now recognized that non-linear coupling in the motion of accreting
fluid may be responsible for some of the observed features of (kHz) QPOs
in neutron-star and black-hole systems. It has been suggested that a
non-linear resonance may be responsible for these twin QPOs 
\citep{ka2,shoji,tokyo}. Among the reasons to believe that these highest
frequencies reflect a non-linear resonance between two oscillation modes
of a disk, probably occurring only in strong-field gravity,  are the
$3:2$ ratio of the frequencies of the twin QPOs in black holes
\citep{ak01,mcr},  and the sub-harmonic frequency difference between the
two QPOs observed in the accreting $2.5$~ms pulsar
\citep{rudy,ketal,lak}. It has been found that a parametric-like
resonance can be excited between the radial and vertical epicyclic
frequencies in a simplified mathematical model of fluid motion in
Einstein's gravity, and that the ratio of the two kHz frequencies can be
made to match closely the one observed in Sco X-1 if the source is
slightly off-resonance \citep{akklr03,paola}. In this paper, we employ
the same model of twin kHz QPOs and we investigate its consequences for
the time behavior of predicted frequencies and amplitudes. We lay down
our description within the framework of an (internal) resonance model
describing the oscillatory modes of a conservative system. Details of the
adopted method and calculations can be found elsewhere \citep{h04}.

\section{The model}
We solved a very general system of two coupled nonlinear oscillators
described by the following governing  equations near the resonance
$\omega_\theta:\omega_r \approx 3:2$,
\begin{eqnarray}
\ddot{\delta r} + \omega_r^2 \,\delta r &=
& \omega_r^2 \,f(\delta r, \delta\theta, \dot{\delta r},
\dot{\delta \theta}), 
\label{eq1}\\
\ddot{\delta \theta} + \omega_\theta^2 \,\delta \theta &=
& \omega_\theta^2 \,g(\delta r, 
\delta\theta, \dot{\delta r}, \dot{\delta \theta}),
\label{eq2}
\end{eqnarray}
under the condition that these two equations are invariant with respect to
reflection of time, i.e., the Taylor expansion of functions $f$ and $g$
does not contain odd powers of the time derivatives of $\delta{r}$ and 
$\delta\theta$. These equations may be taken to approximate  two modes of an
accretion disk with eigenfrequencies close to the radial and vertical
epicyclic frequencies, $ \omega_r$ and $\omega_\theta$, respectively. 
The functions $f$, $g$ depend on the space-time metric, as well as on
the properties of the fluid flow. Any particular mechanism should specify 
the physical meaning of these
functions and fix their values, e.g.\ in terms of the accretion
rate, resonance radius and other parameters. Such special models 
have been indeed proposed and they are encompassed by the
scheme developed herein. For example, a special form of $f$ and $g$ was 
assumed in \citet{akklr03} and \citet{paola}, based on the idea of
parametric resonance, but here we keep the 
discussion general and use the special solution for the purpose of an 
example and a numerical check, as described below. In fact, from the 
mathematical point of view, the 
free parameters of the model are the expansion
coefficients through the fourth order of the functions $f$ and $g$,
which unambiguously define the solution.

We applied the method of multiple time scales \citep{nm79}
and looked for a solution in the form of a series,
\begin{eqnarray}
\delta r(t) &=& \epsilon r_1 + \epsilon^2 r_2 + \epsilon^3 r_3 + \epsilon^4 r_4
 +{\cal O}(\epsilon^5),\\
\delta \theta(t) &=& \epsilon\theta_1 + \epsilon^2\theta_2 + \epsilon^3\theta_3 
 + \epsilon^4\theta_4 +{\cal O}(\epsilon^5).
\end{eqnarray}
This
approach enables us to find a uniformly converging solution, which
does not suffer from secularly growing terms. We
introduced new independent variables $T_k \equiv \epsilon^k\,t$ and
derived the conditions for the elimination of secular terms in the
``fast'' variable  $T_0=t$, up to the fourth order in the expansion
parameter $\epsilon\ll1$. We restrict ourselves to the leading terms
of the expansions, which take the form
\begin{eqnarray}
\delta r(t)&=&A_r (t)\, e^{i\omega_r t} + {\rm c.c.},
\label{eq3} \\
\delta \theta(t)&=&A_\theta (t)\, e^{i\omega_\theta t} + {\rm c.c.}.
\label{eq4}
\end{eqnarray}
The solvability conditions describe the time evolution of
complex amplitudes, which we further rewrite in the form
\begin{equation}
A_r \equiv \frac{1}{2} a_r\, e^{i\phi_r},\quad
A_\theta \equiv \frac{1}{2} a_\theta\, e^{i\phi_\theta}.
\end{equation}
The amplitudes and phases therefore satisfy equations
\begin{eqnarray}
\dot{a}_r &=& \frac{\alpha\omega_r}{16}\, a_r^2\,a_\theta^2\,\sin \gamma, 
\label{eq:ar} \\
\dot{a}_\theta &=
& - \frac{\beta \omega_\theta}{16}\,a_r^3\,a_\theta\,\sin \gamma, 
\label{eq:atheta} \\
\dot{\phi}_r &=& - \frac{\omega_r}{2}\, 
\left[\kappa_r\,a_r^2 + \kappa_\theta\,a_\theta^2 \right] -
\frac{\alpha \omega_r}{16}\, a_r\,a_\theta^2\,\cos \gamma,
\label{eq:phir} \\
\dot{\phi}_\theta &=
& - \frac{\omega_\theta}{2}\, \left[\lambda_r\,a_r^2 + \lambda_\theta\,
a_\theta^2 \right] - \frac{\beta \omega_\theta}{16} a_r^3\,\cos \gamma, 
\label{eq:phitheta}
\end{eqnarray}
where $ \gamma \equiv 2 \phi_\theta - 3 \phi_r - \sigma t $ and $ \sigma
=  3 \omega_r -  2 \omega_\theta $ are a phase function and a detuning
parameter, and  $ \alpha $, $ \beta $,  $ \kappa_r $, $ \kappa_\theta $,
$ \lambda_r $ and $ \lambda_\theta $  are real constants describing our
system. The detuning parameter $\sigma$ describes small departures of 
the eigenfrequencies from sharp rational ratio.
In this way, the method of multiple scales allows us to capture
the amplitude--frequency interaction that is present in the oscillating
solution.

Equations (\ref{eq:phir}) and (\ref{eq:phitheta}) provide frequency 
corrections $\Delta \omega_r = \dot{\phi}_r$ and  $\Delta
\omega_\theta = \dot{\phi}_\theta$ to the eigenfrequencies  $\omega_r$ 
and $\omega_\theta$. The actual,
corrected frequencies will be denoted by an asterisk in order to
distinguish them from the corresponding eigenfrequencies:
$\nu_r^\ast\equiv(\omega_r+\Delta\omega_r)/(2\pi)$,
$\nu_\theta^\ast\equiv(\omega_\theta+\Delta\omega_\theta)/(2\pi)$.
Mathematical validity of the adopted approximation
requires the frequency corrections to be small in our model. 

\begin{figure}
\begin{center}
\FigureFile(0.48\textwidth,0.48\textwidth){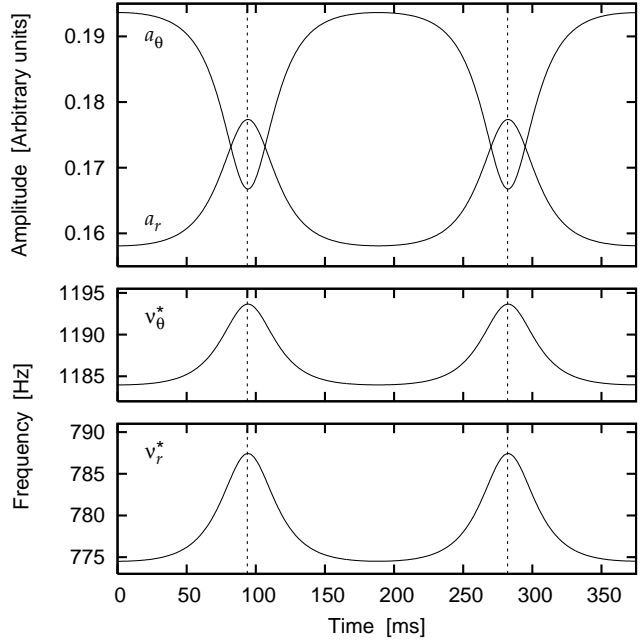}
\end{center}
\caption{Time evolution of amplitudes (top panel) and the epicyclic
frequencies (vertical in the middle, radial at bottom) of high-frequency 
QPOs. The two epicyclic modes are interpreted as radial/vertical 
oscillations of the accreting fluid. The variation of the oscillation
amplitudes is {\it not} arbitrary in the adopted resonance model:
instead, a relation between the amplitudes and frequencies is naturally
predicted.}
\label{fig:fig1}
\end{figure}

Equations (\ref{eq:ar}) and (\ref{eq:atheta}) can be solved for 
$\sin\gamma$ and integrated to read
\begin{equation}
\label{eq:energy}
a_r^2 + \frac{\alpha\omega_r}{\beta\omega_\theta}\,a_\theta^2 \equiv E = 
{\rm const}.
\end{equation}
The quantity $E$ is proportional to the energy of oscillations 
and is conserved in this order of approximation. Equation~(\ref{eq:energy})
describes an ellipse in ($a_r,a_\theta$) coordinates and it rules 
the flow of energy between the oscillation modes.

The two epicyclic frequencies $\omega_r$ and $\omega_\theta$ define the
basic modes of oscillations. Secular terms and the possibility of
parametric resonance appear in the second order of approximation. 
In the third order, the frequency corrections are characterized by the 
foursome of parameters: $\kappa_r$, $\kappa_\theta$, $\lambda_r$ and
$\lambda_\theta$. Finally, the two additional parameters, $\alpha$ and
$\beta$, provide the fourth-order corrections, and describe the
semi-axes ratio of the energy ellipse. The above-given
list completes the most general description of any system described
by equations of the form (\ref{eq1})--(\ref{eq2}).

The general solution corresponds to the periodic exchange of 
energy between the two oscillators. If the system is far from 
the steady state, the amplitudes and frequencies of oscillations
fluctuate, maintaining the energy condition (\ref{eq:energy}). The
period of energy exchange is approximately given by 
\begin{equation}
\label{eq:period}
T \sim \frac{16 \pi}{\beta\omega_\theta}\; E^{-3/2}.
\label{eq:ene}
\end{equation}
On the other hand, particular steady-state solutions are possible 
in which the observed frequencies of the two modes remain strictly 
constant in time in $3:2$ ratio, despite the fact that the eigenfrequencies
depart from it. Such solutions play a role of singular points of the system
(\ref{eq:ar})--(\ref{eq:phitheta}). Close to steady state the 
approximation (\ref{eq:ene}) fails and the period becomes greater 
\citep{h04}.

In order to verify accuracy of our work, we checked that the numerical
solution of a special form  \citep{akklr03} of the system of equations
(\ref{eq1})--(\ref{eq2}), in which the right-hand side corresponded to
the case of nearly geodesic circular motion in the pseudo-Newtonian
potential of Paczy\'nski and Wiita, does indeed closely follow the 
ellipse of eq.~(\ref{eq:energy}). Therefore we can be confident
that the above-described analytical method gives credible
results.

It is evident from eqs.\ (\ref{eq:ar}), (\ref{eq:atheta}),
(\ref{eq:energy}), and (\ref{eq:ene}), that the amplitudes and
frequencies of high-frequency QPOs are modulated at the frequency
$\nu_3\equiv1/T$, which is related to the amplitudes  of the original
high-frequency QPOs. We remind the reader that here we 
restrict ourselves to the case $\omega_\theta:\omega_r=3:2$, 
however, similar results can be
obtained also for other eigenfrequency ratios.

\section{Low-frequency modulation}
In the general discussion here, our model has six free parameters,
corresponding to (a combination of) the lowest order expansion
coefficients in the Taylor series of the unspecified functions $f$ and
$g$ in equations  (\ref{eq1}) and (\ref{eq2}). Hence, we cannot predict
unique behavior.  However, it is remarkable that a modulation of
frequencies and amplitudes follows naturally from the assumptions made.
Because the oscillators are non-linear, their frequency varies with
amplitude. Because the two oscillators are coupled in a system with
constant energy, the amplitudes of the oscillators are anticorrelated.
Because  the two oscillators are in resonance, their frequencies are
correlated.

It is not lost on us that similar correlations --- discussed by
\citet{Yu} --- hold also on long timescales. For example, 
\citet{sco} show that as the kHz frequencies increase, the ratio of
power in the upper to lower kHz QPOs decreases (up to a point). Our
Figure~1 demonstrates this to be the case also here. We note that in
previous work, a particular relation was found to hold between the two
frequencies of the system, in agreement with that observed in long-term
variations of the twin kHz QPO frequencies in Sco X-1 
\citep{akklr03,paola}. Here, we consider variations on shorter
timescales.

As we mentioned above [see equation~(\ref{eq:ene})], the timescale of
modulation in our model is directly related to energy $E$, i.e., the
weighted sum of amplitudes squared, which we adjusted to obtain a
modulation at the NBO frequency of Sco X-1, $\nu_3\approx 6\,$Hz. We
identify the corrected frequencies of our model, i.e., $\nu_r^*$ and
$\nu_\theta^*$, with the twin kHz QPO frequencies $\nu_{\rm{}lower}$ and
$\nu_{\rm{}upper}$, respectively. 

Figure~1 (top panel) exhibits the time variation of the ``radial'' and 
``vertical'' amplitudes of oscillations. In the lower panels we show the
correlation between the two kHz frequencies found in our solution. By
the assumption of nearly $3:2$ resonance, the two frequencies of actual
oscillations satisfy relation $\nu^*_{\rm upper}\approx 1.5\nu^*_{\rm
lower}$. Notice a perfect correlation between the variation of the lower
amplitude and the variation of the upper frequency $\nu^*_{\rm upper}$.
This we interpret as the same correlation that was found by \citet{Yu};
compare their figure~2. The magnitude of frequency variation agrees to
within a factor of $3$ with the data: it is $20$~Hz for
$\nu_{\rm{}upper}=1.1\,$kHz in the data, and about $7$ parts in
$1000$~in our calculation. 

\section{Discussion} 
We have found that a non-linear resonance in a system simulating an
accretion disk, invoked previously to explain the appearance of two
frequencies in approximately $3:2$ ratio in black-hole and  neutron-star
X-ray data, results in a periodic time variation of the frequencies. 

The Fourier transform of the frequency-modulated and amplitude-modulated
periodic signal would result in several sidebands, in practice leading
to an increase in width of the (noisy) signal. If the observed radiation
flux were modulated with the squared modulus of the amplitudes
(\ref{eq3})--(\ref{eq4}), its Fourier transform would exhibit a (weak)
component at low frequency of the modulation apparent in figure~1, in
addition to the high-frequency signal with its sidebands. In this
exploratory work, we are not attempting to model the full power spectrum
of QPO sources and, as yet, we have made no attempt to translate the
amplitudes of motion in the model into modulations of the X-ray flux.
Strictly speaking, our toy-model gives a coherent signal rather than a
QPO, and no details of the excitation or damping were modelled.

It has been suggested that the high-frequency QPOs vary on a timescale
of seconds in some sources, notably in Sco X-1 (\cite{Yu}) and in the
black hole candidate XTE J1550-564 (\cite{Yu2}). The low-frequency
modulation occurs at $\approx6$~Hz in both sources. In the above
described calculation we were able to reproduce this variation for Sco
X-1, including the anti-correlation between the amplitude of the lower
peak and the frequency of the upper one. A similar approach is possible
also in case of XTE J1550-564, although the black hole candidates
typically exhibit lower frequencies compared to those in neutron stars,
and particularly to those in Sco X-1. This means that if the $6$~Hz QPO
seen by \cite{Yu2} corresponds to the modulation discussed in our model,
then the ratio $\nu_{\rm{}lower}/\nu_3$ has to be set differently (about
30 in XTE 1550-564 with $\nu_{\rm{}lower}=184$~Hz). And this in turn
implies $E$ is different in both systems.

If the correspondence of our results with the observed modulation of kHz
QPO properties on the $6$~Hz NBO timescale is not accidental, for the
first time we would have a physical explanation for the presence of the
rather low frequency QPO in what is otherwise a domain of rapid
variability. The larger point is that non-linear resonance (most likely
between modes of oscillation possible only in strong gravity) holds
promise for explaining not only the highest frequencies observed in
accreting neutron stars and black holes, but also the mysterious
phenomenology of low frequency features in the power density spectrum,
without invoking additional mechanisms.

\medskip
It is a pleasure to acknowledge the hospitality of the Director and
staff of UK Astrophysical Fluids Facility at Leicester University, where
this work was begun, and of Sir Franciszek Oborski, the master of
Wojnowice Castle, where it was completed. Research supported in part by
the European Commission  {\it Access to Research Infrastructure Action
of the Improving Human Potential Program} at the UKAFF, by KBN (grant
2P03D01424), Charles University (grant 299/2004) and by the Swedish
Research Council. JH and VK are grateful for support from the Czech
Science Foundation via grants 205/03/H144 and 205/03/0902.

\end{document}